\documentclass[12pt]{iopart}
\usepackage{iopams, graphicx, setstack,xcolor}

\begin{document}

\title[Dissipation at limited resolutions]{Dissipation at limited resolutions: Power law and detection of hidden dissipative scales}

\author{Qiwei Yu$^{1,*}$ and Pedro E. Harunari$^{2,*}$}

\address{$^1$ Lewis-Sigler Institute for Integrative Genomics, Princeton University, Princeton, NJ 08544}
\address{$^2$ Complex Systems and Statistical Mechanics, Department of Physics and Materials Science, University of Luxembourg, L-1511 Luxembourg City, Luxembourg}
\address{$^*$ These two authors contributed equally to this work.}

\eads{\mailto{qiweiyu@princeton.edu}, \mailto{pedro.harunari@uni.lu}}

\vspace{10pt}

\begin{abstract}
Nonequilibrium systems, in particular living organisms, are maintained by irreversible transformations of energy that drive diverse functions. Quantifying their irreversibility, as measured by energy dissipation, is essential for understanding the underlying mechanisms. However, existing techniques usually overlook experimental limitations, either by assuming full information or by employing a coarse-graining method that requires knowledge of the structure behind hidden degrees of freedom. Here, we study the inference of dissipation from finite-resolution measurements by employing a recently developed model-free estimator that considers both the sequence of coarse-grained transitions and the waiting time distributions: $\sigma_2=\sigma_2^\ell + \sigma_2^t$. The dominant term $\sigma_2^\ell$ originates from the sequence of observed transitions; we find that it scales with resolution following a power law. Comparing the scaling exponent with a previous estimator highlights the importance of accounting for flux correlations at lower resolutions. $\sigma_2^t$ comes from asymmetries in waiting time distributions. It is non-monotonic in resolution, with its peak position revealing characteristic scales of the underlying dissipative process, consistent with observations in the actomyosin cortex of starfish oocytes. Alternatively, the characteristic scale can be detected in a crossover of the scaling of $\sigma_2^\ell$. This provides a novel perspective for extracting otherwise hidden characteristic dissipative scales directly from dissipation measurements. We illustrate these results in biochemical models as well as complex networks. Overall, this study highlights the significance of resolution considerations in nonequilibrium systems, providing insights into the interplay between experimental resolution, entropy production, and underlying complexity.
\end{abstract}

\vspace{2pc}

\submitto{\JSTAT}
\maketitle

%
%

\section{Introduction}

Although the pursuit of ever-increasing resolution is a primary goal of technological progress, there are many instances where coarser observations enhance features that would otherwise be missed. A technique known as image binning lumps adjacent pixels to increase signal-to-noise ratio at the expense of resolution, and it has been an important tool in the discovery of faint objects in the universe \cite{liSpatialMetallicityDistribution2023, sandersContourBinningNew2006}. Regarding emergent phenomena, the microscopic interactions between individual components often do not suffice to appreciate complex large-scale phenomena, such as the chemical reactions in Turing patterns \cite{turingChemicalBasisMorphogenesis1990} or the rules for Conway's game of life \cite{gardnerMathematicalGames1970}. In these cases, improving measurement resolution might introduce additional computational requirements without aiding the detection of patterns. In a more routine experience, squinting the eyes can reveal a figure in optical illusions such as hybrid images \cite{olivaHybridImages2006}, with the Monroe/Einstein image being the most famous example. Across resolutions, observables present different behaviors and correlations, and hidden structures can be uncovered if the right observable is measured. In this contribution, we demonstrate that for nonequilibrium systems, quantifying energy dissipation across scales reveals new insights into the underlying structures.

The vast majority of biological phenomena are intrinsically nonequilibrium processes sustained by the continuous dissipation of free energy~\cite{yang_physical_2021}, whose quantification is key for understanding the dynamics and energetics of biological function and physical processes. Examples of these processes are adaptation~\cite{hazelbauer_bacterial_2008, lan_energyspeedaccuracy_2012}, error correction~\cite{hopfield_kinetic_1974,ninio_kinetic_1975,murugan_speed_2012,yu_energy_2022}, and environment sensing \cite{prigogine1987exploring, nicoletti2023adaptation, ouldridge2017thermodynamics, mehta2016landauer, theWolde2016fundamental, govern2014optimal, hathcock_nonequilibrium_2023,tjalma_trade-offs_2023}. Multiple temporal and spatial scales can be involved: While free energy is often harnessed on the molecular scale (e.g. from the hydrolysis of energy-rich molecules such as ATP), it can be used to drive processes at much larger scales, like pattern formation~\cite{wigbers_hierarchy_2021} and collective motion~\cite{vicsek_novel_1995,toner_long-range_1995, toner_flocks_1998, ferretti_signatures_2022, yu_flocking_2022}. Such multiscale nonequilibrium processes can be studied in model experimental systems such as the actomyosin cortex of a starfish oocyte~\cite{tan_scale-dependent_2021} and microtubule active gels~\cite{foster_dissipation_2023}. Understanding how much free energy is dissipated on those different scales can lead to mechanistic insights into the intricacies of the underlying structure, such as the characteristic timescale of active processes~\cite{tan_scale-dependent_2021}. However, this is usually difficult due to limitations on the scales and degrees of freedom that can be resolved in experiments. It is crucial to elucidate how much information can be extracted from measurements with finite resolution.

From a theoretical perspective, the thermodynamics of coarse-grained systems have been studied in distinct scenarios, where the most prominent measure of dissipation is the entropy production rate (EPR). 
Previous studies have considered different forms of coarse-graining:
Timescale separation~\cite{espositoStochasticThermodynamicsCoarse2012, rahavFluctuationRelationsCoarsegraining2007, kawaguchiFluctuationTheoremHidden2013, boEntropyProductionStochastic2014, boMultiplescaleStochasticProcesses2017, baiesiEffectiveEstimationEntropy2023} represents the possibility of monitoring slow degrees of freedom while fast ones go undetected; decimation~\cite{martinezInferringBrokenDetailed2019} considers subsets of states and transitions as observables and can preserve the full entropy production~\cite{teza_exact_2020}; milestoning has been used to map continuous dynamics onto the framework of discrete state space and ensures thermodynamic consistency~\cite{hartichEmergentMemoryKinetic2021, hartichViolationLocalDetailed2023}; lumping refers to merging states that cannot be resolved due to, e.g., proximity in space, and often leads to a drastic decrease in EPR at the coarse-grained level~\cite{yu_inverse_2021,yu_state-space_2022}. 
Forms of coarse-graining can also be inspired by basins of attraction \cite{falascoLocalDetailedBalance2021, falascoMacroscopicStochasticThermodynamics2023}, first-order phase transitions \cite{nguyenExponentialVolumeDependence2020, fioreCurrentFluctuationsNonequilibrium2021}, and imperfect measurements \cite{ferri-cortesEntropyProductionFluctuation2023, borrelliFluctuationRelationsDriven2015}. Obtaining the entropy production or finding its upper/lower bounds provides key insights into the system not only because it estimates the real EPR or establishes the minimal thermodynamic cost of a process, but it also establishes bounds for efficiency~\cite{leightonInferringSubsystemEfficiencies2023, pietzonka2016universal, Dechant2018}. Although much progress has been made to extract the EPR from statistics of the coarse-grained dynamics, e.g. using waiting time distributions~\cite{harunari_what_2022, van_der_meer_thermodynamic_2022, vandermeerTimeResolvedStatisticsSnippets2023, martinez2019inferring, berezhkovskii_forwardbackward_2019, blom_milestoning_2023, PhysRevLett.105.150607, skinner_estimating_2021, kapustinUtilizingTimeseriesMeasurements2024, nitzanUniversalBoundsEntropy2023, ertelEstimatorEntropyProduction2023, harunari2024unveiling}, it remains an open question how much can be learned about the microscopic system from coarse-grained observations.

Here, we explore the scenario of measurement with limited resolutions. These are typically experiments where not all fine-grained degrees of freedom are distinguished by the measurement apparatus, resulting in unresolved trajectories. To represent this scenario, we coarse-grain the system by lumping states that are sufficiently similar, i.e., close in terms of a relevant distance measure. This allows us to examine the effect of varying resolution by changing the distance threshold for lumping states, which reveals the dependence of detected dissipation with resolution. Notice that decreasing resolution can always be done in the post-processing, hence the data can be treated to unveil the properties we discuss without the need for always improving resolution.

Measuring EPR in coarse-grained systems is generally difficult because it depends on the statistics of current observables which, after coarse-graining, do not share simple relations with fully resolved currents~\cite{ptaszynskiFirstpassageTimesRenewal2018}.
Previously, it was shown that the apparent entropy production rate at coarse-grained scales decreases following an inverse power law, with an exponent that depends both on the topology of the state space and the correlation of the probability fluxes~\cite{yu_inverse_2021,yu_state-space_2022}. When the fluxes are negatively correlated (i.e. frequent back-and-forth transitions), the apparent EPR decreases faster than the number of coarse-grained transitions. This suggests that harnessing the information encoded by flux correlations might result in a more accurate estimation of the EPR.
Indeed, recent works have made significant progress in estimating the EPR of systems with partially visible transitions through specialized estimators that take into account the correlation (through sequence and waiting time) between coarse-grained transitions~\cite{harunari_what_2022, van_der_meer_thermodynamic_2022,ertelEstimatorEntropyProduction2023, harunari2024unveiling}. Hence, it is natural to ask whether applying this approach to coarse-grained measurements can reveal more information. We focus on one estimator that is obtained by the sum of two contributions, one from the sequence of visible transitions and one from their waiting times; they are affected in distinct ways by changes in resolution and will prove to have different roles in the quantitative assessment of dissipation and internal scales. Importantly, the estimator strictly bounds the EPR from below. 

When applying the specialized estimator to limited-resolution measurements, we find that the estimator leads to more accurate apparent values of EPR and provides mechanistic insights into the dissipative scales. First, we show that the apparent EPR estimated from this approach decreases with the coarse-grained scale following a power law, with an exponent that is smaller than that of the direct coarse-graining approach. Thus, accounting for flux correlation drastically improves EPR estimation.  
In addition, we show that the irreversibility from the waiting time distribution follows a non-monotonic relation with the coarse-grained scale, with its peak position reflecting the dissipative scale of the system. 
This is similar to a non-monotonic relation reported in the actomyosin cortex~\cite{tan_scale-dependent_2021}, where the peak position corresponds to the dissipative timescale, and a spatial counterpart to the detection of dissipative timescales through temporal coarse-graining \cite{cisnerosDissipativeTimescalesCoarsegraining2023}.
If multiple scaling regimes are present, their crossover may be detected by a crossover in the EPR scaling. 
These results can be readily applied to experimental data, as illustrated in biochemical reaction systems such as Brusselator~\cite{nicolis_self-organization_1977, qian_concentration_2002, fritz_stochastic_2020} and Schl\"ogl models~\cite{schlogl_chemical_1972,vellela_stochastic_2009}. 
We also find similar relations in state networks with non-regular topologies~\cite{ER1959,Watts1998,barabasi_emergence_1999,barabasi_scale-free_2009}, which may be useful for analyzing time irreversibility in complex networks~\cite{jung_entropy_2020, PhysRevE.100.012104} such as neural networks present in brain dynamics~\cite{sanzperlNonequilibriumBrainDynamics2021, lynn_broken_2021, lynn_decomposing_2022}.

%
%

\section{Formalism}

We consider a system whose dynamics is described by a continuous-time Markov chain among discrete mesoscopic states. These states capture configurations that are of thermodynamic relevance, and the transition rates between them include the influence of the environment. Its EPR, quantifier of the statistical asymmetry between a process and its time-reversal, is given by
\begin{eqnarray}\label{eq:realEPR}
    \sigma = \mathcal{K}\sum_{\ell} P(\ell) \ln \frac{P( \ell)}{P(\bar{\ell})},
\end{eqnarray}
where $\ell$ sums over all transitions with $\bar{\ell}$ being its reverse. $P(\ell)$ is the probability of observing transition $\ell$, and $\mathcal{K}$ is the dynamical activity defined as the average number of transitions per unit time. 
We start with a microscopic description where all the transitions are visible and the steady-state probability $P(\ell)$ can be obtained by solving the master equation. After coarse-graining, only a subset of transitions remain observable, leading to an observed activity $\mathcal{K}_{\rm{obs}}<\mathcal{K}$.

\begin{figure*}[t]
    \centering
    \includegraphics[width=\textwidth]{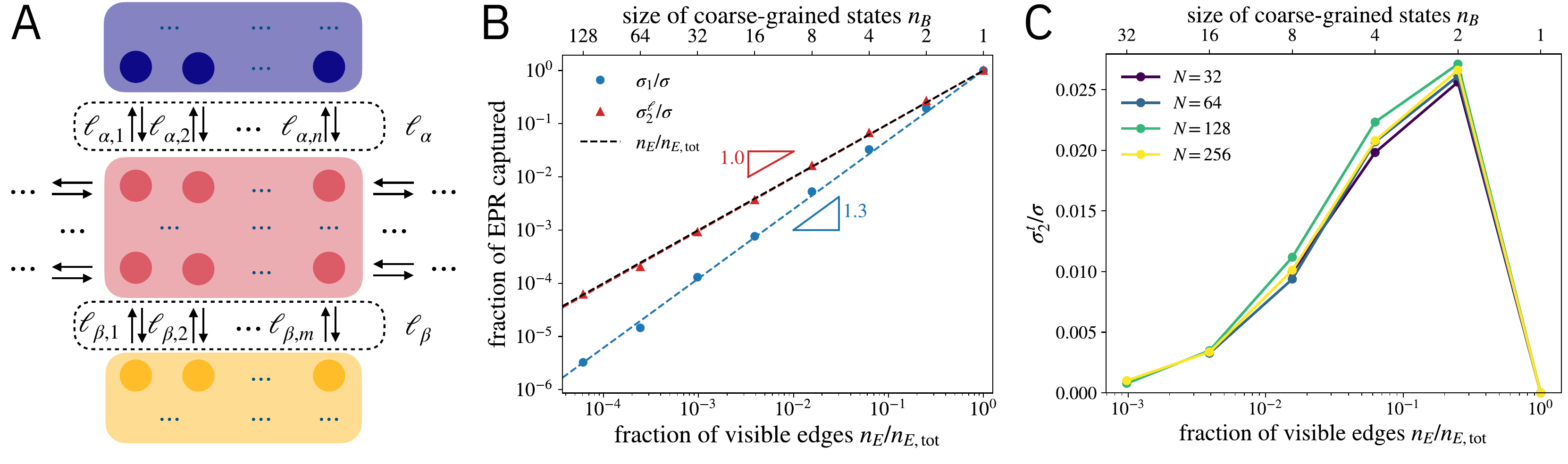}
    \caption{(A) Schematics of the theoretical framework for coarse-graining both states and transitions. Circles and shaded boxes label microscopic and coarse-grained states, respectively. Arrows represent microscopic transitions that will be combined into coarse-grained transitions (boxes with dashed lines $\ell_{\alpha}$, $\ell_\beta$). 
    (B--C) Fraction of the entropy production captured by coarse-grained measurements via estimators based on (B) the sequence of observable transitions and (C) the waiting time distributions. In both panels, the model considered is a square lattice with rates drawn from a standard lognormal distribution. The black dashed line represents a trivial scaling relation in which the average observed EPR per transition remains constant.
    Panel (B) starts with a $1024\times 1024$ state space at the fine-grained level, while panel (C) has lattices of different sizes $N^2$.
    }
    \label{fig:1}
\end{figure*}

To investigate how limited resolution affects dissipation, we adopt a coarse-graining procedure that lumps together states and transitions identified by sufficiently similar degrees of freedom.
For illustrative purposes, we start with a square lattice where states are identified by two degrees of freedom, which can be, for instance, spatial positions or chemical concentrations (see \fref{fig:1}A).
The proposed procedure representing limited resolutions joins in a coarse-grained state (shaded boxes) ``microscopic states'' belonging to a neighborhood of a given size. All microscopic states within the same coarse-grained state cannot be resolved, and thus transitions between them become invisible. Furthermore, all ``microscopic'' transitions $\ell_{\alpha,i}$ that stem from one coarse-grained state to another are observed as the same coarse-grained transition $\ell_\alpha$.
A typical measurement yields a sequence of coarse-grained transitions and the waiting times between consecutive transitions (intertransition times): $(\ell_\alpha, t_\alpha; \ell_\beta, t_\beta; \ldots)$. The probability of observing transition $\ell_\alpha$ and, after time $t$, $\ell_\beta$, is given by the sum of the probability of its constituent microscopic transitions
\begin{eqnarray}\label{eq:cg_joint}
    P(\ell_\alpha, \ell_\beta ;t) = \sum_{i,j} P(\ell_{\alpha,i}, \ell_{\beta,j};t).
\end{eqnarray}
Similarly, the joint probability of a sequence of transitions is given by the marginalization
\begin{eqnarray}\label{eq:cg_marginal}
    P(\ell_\alpha, \ell_\beta)=\int_0^{\infty} P(\ell_\alpha, \ell_\beta ;t) \rmd t = \sum_{i,j} P(\ell_{\alpha,i}, \ell_{\beta,j}).
\end{eqnarray}

While the low-resolution dynamics is typically non-Markovian, which makes it difficult to estimate the true EPR, the experimenter can still compute the apparent (or ``local'') EPR ~\cite{lynn_decomposing_2022,lynn_emergence_2022} from the statistics of the observed coarse-grained transitions. The simplest estimator accounts for the statistics of each coarse-grained transition individually:
\begin{equation}\label{eq:sigma_0}
    \sigma_1 =  \mathcal{K}_{\rm{obs}}\sum_{\alpha} P(\ell_\alpha) \ln \frac{P( \ell_\alpha)}{P(\bar{\ell}_\alpha)},
\end{equation}
where $\ell_\alpha$ sums over coarse-grained transitions. This quantity is readily available from empirical data since the probabilities can be estimated from the frequencies of observable transitions. It bounds the full EPR from below by only considering the absolute probabilities of each transition, disregarding hidden degrees of freedom, the presence of limited resolutions, and the fact that trajectories are non-Markovian.

Higher-order statistics can be included to build better estimators by also considering correlations between distinct transitions, usually in the form of joint probabilities. Transition-based estimators that consider the statistics of pairs of transitions and the waiting time between them have recently been developed \cite{harunari_what_2022, van_der_meer_thermodynamic_2022}. 
However, it is unclear whether they rigorously bound EPR when observables are lumped transitions consisting of many indistinguishable transitions. 
To overcome this, we consider an alternative second-order estimator that strictly bounds EPR from below when applied to lumped transitions~\cite{harunari2024unveiling, ertelEstimatorEntropyProduction2023}. The estimator can be split into sequence and waiting time contributions: $\sigma_2 \equiv \sigma_2^\ell + \sigma_2^t$, where
\begin{eqnarray}\label{eq:sigma_ell}
    \sigma_2^\ell = \frac{\mathcal{K}_{\rm{obs}}}{2} \sum_{\alpha, \beta} P(\ell_\alpha, \ell_\beta) \ln \frac{P( \ell_\alpha, \ell_\beta )}{P( \bar{\ell}_\beta , \bar{\ell}_\alpha)},
\end{eqnarray}
and
\begin{eqnarray}\label{eq:sigma_t}
    \sigma_2^t = \frac{\mathcal{K}_{\rm{obs}}}{2} \sum_{\alpha, \beta} P(\ell_\alpha, \ell_\beta) D_{\rm{KL}}[P(t|\ell_\alpha, \ell_\beta)||P(t|\bar{\ell}_\beta,\bar{\ell}_\alpha)],
\end{eqnarray}
with $D_{\rm{KL}}$ denoting the Kullback-Leibler divergence between waiting time distributions. Importantly, $\sigma_2$ estimates the EPR by forming a lower bound to the EPR that is tighter than that of $\sigma_1$ because it retains more information on the underlying dissipative processes. Notice that this estimator is agnostic to the experimental resolution, it can be applied even when it is not known whether the experiment is able to capture all transitions.

For applications, the probabilities of transitions $\ell$ and their waiting times can be directly extracted from experimental measurements, with $\ell$ representing all transitions between distinguishable states at a given resolution. 
Here, we illustrate the usage of the estimator $\sigma_2$ in model systems, where we use the survival operator approach (see Appendix A for details) to evaluate the joint distribution of transitions and waiting times.

In the following, we observe that the two components of the more specialized estimator, $\sigma_2$, play different roles. 
Section~\ref{subsec:scaling} studies $\sigma_2^\ell$, which tends to be larger and more important for estimating the total EPR; we find that it scales with resolution following a power law.
Section~\ref{subsec:scales} focuses on the detection of internal scales through the distinct exponents of $\sigma_2^\ell$ and the peak of $\sigma_2^t$; both provide information on the internal structure such as the characteristic temporal and spatial scales of dissipation. 
Section~\ref{sec:illustrations} illustrates all of these findings in two standard biochemical models, a biochemical oscillator (the Brusselator model) and a system with multiple stable states (the Schl\"ogl model).  
We also extends the results to networks of complex topologies, where coarse-graining can be done either by lumping in an embedded space or by random selection of visible transitions. 
Finally, Section~\ref{sec:discussion} further discusses applications to experimental data and future directions. 

%
%

\section{Properties of dissipation at different resolutions}\label{sec:properties}

The estimator $\sigma_2$ measures the apparent dissipation at a given resolution, unraveling it into contributions from sequences of transitions and from waiting times reveals their distinct behavior in terms of resolution changes. In this section, we explore two main features of the estimator, namely the presence of scaling and the prospect of detecting otherwise hidden scales, and their illustrations are provided in the following section.

\subsection{Scaling of $\sigma_2$}\label{subsec:scaling}

For simplicity, we cast the arguments herein for a square lattice (\fref{fig:1}A) but they naturally extend to different topologies. Coarse-graining is done by merging $n_{\rm{B}}$-by-$n_{\rm{B}}$ square blocks into a single (coarse-grained) state; intrablockrablock transitions become hidden, and interblock transitions are lumped into $n_{\rm{E}}$ visible transitions. Each interblock transition $\ell_{\alpha}$ is composed of $n_{\rm{B}}$ microscopic transitions $\ell_{\alpha,i}$ which cannot be resolved in measurements. As $n_{\rm{B}}$ increases, the number of visible transitions decreases as $n_{\rm{E}}\propto n_{\rm{B}}^{-2}$, causing a scaling in the number of terms of $\sigma_1$ and $\sigma_2$. The probabilities involved in these expressions scale with $n_{\rm{B}}$ in a nontrivial manner partly due to the correlations between fluxes. Reference~\cite{yu_inverse_2021} showed that $\sigma_1$ decreases following a power law in terms of block size due to the statistical properties of the lumped fluxes represented by $P(\ell_\alpha)$. A similar argument can be made to show that $\sigma_2^\ell$ also scales with the block size following a power law, where joint probabilities of transition pairs, $P(\ell_\alpha , \ell_\beta)$, play the role of $P(\ell_\alpha)$ [see Eqs.~\eref{eq:sigma_0} and \eref{eq:sigma_ell}]. 
Indeed, while $P(\ell_\alpha)$ represents steady-state fluxes in the state space, $P(\ell_\alpha , \ell_\beta)$ plays the role of fluxes in the transition space, which is the space spanned by all possible transitions of the system (or equivalently, by pairs of consecutive states). As resolution decreases, lumping in the transition space is analogous to lumping in the state space, except that the dimension of the transition space is much higher. Hence, we expect $\sigma_2^\ell$ to decrease following a power law, but the scaling exponent may differ from that of $\sigma_1$ since $P(\ell_\alpha , \ell_\beta)$ follows a different statistical structure than $P(\ell_\alpha)$. 

As shown in \fref{fig:1}B, for a system with i.i.d.~random rates, both $\sigma_1$ (blue) and $\sigma_2^\ell$ (red) decrease following power laws with $n_{\rm{E}}$, consistent with the prediction above. $\sigma_2^\ell$ in general has a smaller exponent than $\sigma_1$, which means that it is not only larger, but its relevance rapidly grows at smaller resolutions. Thus, by accounting for joint probabilities between consecutive coarse-grained transitions, $\sigma_2^\ell$ provides a more accurate estimate of the EPR, which can be orders of magnitude better than $\sigma_1$ for small resolutions (large $n_{\rm{B}}$). The combined estimator $\sigma_2=\sigma_2^\ell+\sigma_2^t$ yields a similar performance (Appendix B) since it is dominated by $\sigma_2^\ell$.

One mechanism behind the different scalings is that the asymmetry $P( \ell_\beta|\ell_\alpha)\neq P( \bar{\ell}_\alpha|\bar{\ell}_\beta)$ captures some of the EPR associated with transitions internal to the coarse-grained state, which is completely discarded in the direct lumping approach ($\sigma_1$). Interestingly, the scaling exponent for $\sigma_2^\ell$ is approximately $1$ (black dashed line), representing a linear scaling with the number of visible transitions, and this same exponent is observed in further examples in the following sections. In other words, the EPR per coarse-grained transition remains constant across resolutions. 

\subsection{Extraction of dissipative scales}\label{subsec:scales}

In systems where the topological and dynamical structures are homogeneous across length scales, such as the square lattice with random rates, the scaling of $\sigma_2^\ell$ persists until coarse-graining approaches the largest scale. This largest scale is the system size for the square lattice, but it can also be the size of a limit cycle or other types of emerging structures. This indicates that, provided that the coarse-graining level does not cross characteristic scales, the dissipation has the same scaling structure across distinct resolutions. However, a system can exhibit drastically different dynamics at different scales (either in state space or in physical real space), which leads to $\sigma_2^\ell$ scaling with distinct exponents for each regime. Therefore, the presence of more than one scaling in $\sigma_2^\ell$ indicates the existence of distinct dissipative scales, with a ``kink'' marking their separation. This will be illustrated in the Schl\"ogl model (see next section).

Another component of the transition-based estimator is $\sigma_2^t$, which captures the irreversibility associated with asymmetric waiting time distributions between coarse-grained transitions, also called intertransition times. $\sigma_2^t$ is always non-monotonic with the block size $n_{\rm{B}}$: at the fine-grained level ($n_{\rm{B}}=1$), the waiting time distribution is exponential and symmetric, leading to $\sigma_2^t=0$; at large $n_{\rm{B}}$, the number of coarse-grained transitions is small, which also leads to small $\sigma_2^t$ (it eventually vanishes when $n_{\rm{B}}$ reaches the system size). 
Thus, $\sigma_2^t$ is maximized at an intermediate scale $n_{B, m}$, where the dynamics within coarse-grained states have maximally asymmetric waiting time distributions upon time reversal. The peak block size $n_{\rm{B,m}}$ provides a natural and model-free measurement of the characteristic scale of the dissipative dynamics.

For square lattice with i.i.d.~random transition rates, $\sigma_2^t$ starts at zero at the finest scale and reaches a maximum at $n_{\rm{B}}=2$ before decreasing monotonically with $n_{\rm{B}}$ (\fref{fig:1}C). Since the rates are drawn independently,  this model does not exhibit long-range structures. Hence,  the lack of a characteristic scale is evidenced by the peak of $\sigma_2^t$ falling at the first level of coarse-graining $n_{\rm{B}} = 2$.
As resolution decreases, the waiting time distributions become increasingly less asymmetric under time reversal, leading to the decrease in $\sigma_2^t$. 
Furthermore, the fraction of observable EPR approximately collapses for different system sizes due to this homogeneity of the rates. When a system has more intricate properties and presents an underlying structure, the peak might reflect the spatial scale of such a structure instead of being localized at the first coarse-graining step. Indeed, the connection between the peaks of irreversibility and internal dissipative timescales will be demonstrated theoretically (e.g. in biochemical oscillators) and was found experimentally in the actomyosin cortex of starfish oocytes~\cite{tan_scale-dependent_2021} (see next section for detailed discussion).

%
%

\section{Illustrations}
\label{sec:illustrations}

\subsection{Brusselator}\label{subsec:brusselator}

\begin{figure*}[t]
    \centering
    \includegraphics[width=0.9\textwidth]{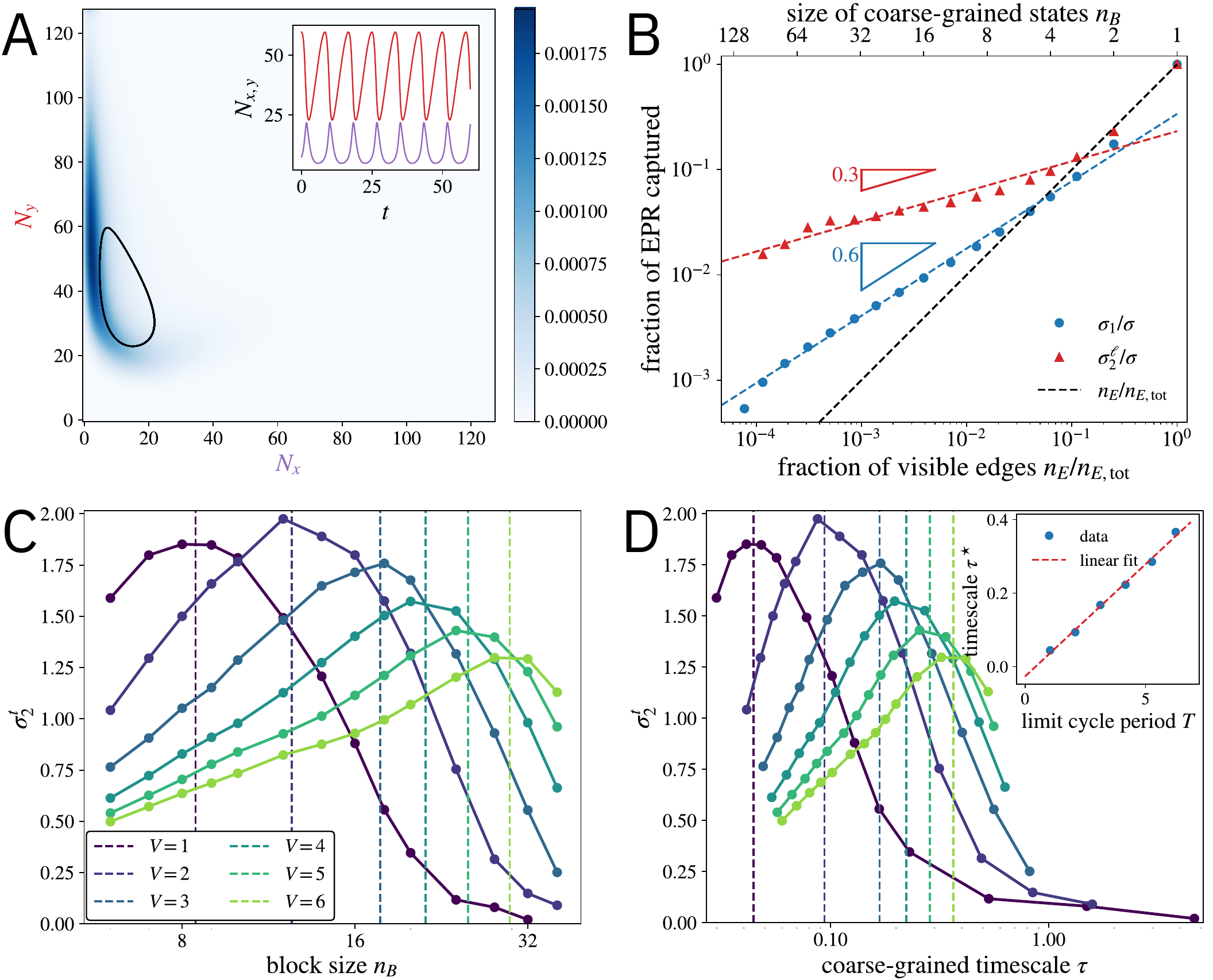}
    \caption{
        Coarse-grained dissipation and characteristic scales in the simplified Brusselator model.
        (A) The colormap represents the steady-state probability distribution, with the limit cycle (black circle) extracted from the oscillation of the deterministic dynamics (inset).
        (B) Fraction of EPR captured ($\sigma_1/\sigma$ and $\sigma_2^\ell/\sigma$) at varying resolutions, with dashed lines approximately representing their scaling regime and the black dashed line representing a scaling of exponent 1.
        (C) $\sigma_2^t$ as a function of the block size $n_{\rm{B}}$ for different volumes $V$, with its maximum indicated by vertical dashed lines.
        (D) $\sigma_2^t$ as a function of the coarse-grained timescale $\tau$, with maxima indicated by vertical dashed lines ($\tau^*$).
        Inset: $\tau^*$ in terms of the period of the limit cycle $T$.
         Parameters: $k_1=5$, $k_{-1}=15$. $k_2=15$, $k_{-2}=0.5$, $k_3=2$, $k_{-3}=1$; $V=60$ for (B). 
    }
    \label{fig:2}
\end{figure*}

To see whether the discussed properties hold in real biochemical systems, we turn to the Brusselator model~\cite{nicolis_self-organization_1977, qian_concentration_2002}, which describes a class of biochemical oscillators. Here, we study the simplified Brusselator model~\cite{fritz_stochastic_2020} to avoid singular behavior at the first coarse-graining iteration in the original Brusselator~\cite{yu_inverse_2021}. The model describes the dynamics of two chemical species $X$ and $Y$ with reactions
\begin{eqnarray} \label{Eq:brusselatorModel}
    A \underset{k_{-1}}{\stackrel{k_{1}}{\rightleftharpoons}} X, \qquad
    B \underset{k_{-2}}{\stackrel{k_{2}}{\rightleftharpoons}} Y, \qquad
    2 X+Y \underset{k_{-3}}{\stackrel{k_{3}}{\rightleftharpoons}} 3 X,
\end{eqnarray}
with $k_{\pm i}$ being kinetic constants of each reaction $\pm i$, and $A$ and $B$ molecules held at constant concentrations. We assume mass-action kinetics, where the transition rates are given by the product of kinetic constants and the concentrations of the substrate; for instance, the forward rate of reaction $3$ is $k_{3}[X]^2 [Y]$, with concentrations $[X]=N_x/V$ and $[Y]=N_y/V$. The state space is a 2D lattice spanned by the number of molecules $N_x$ and $N_y$, with horizontal, vertical, and diagonal transitions corresponding to the three reactions in Eq.~\eref{Eq:brusselatorModel}. In certain parameter regimes, the system exhibits oscillation, represented by a limit cycle in the state space (\fref{fig:2}A). 

Since the transitions within coarse-grained states are highly directional along the limit cycle, we expect $\sigma_2^\ell$ to do a much better job than $\sigma_1$ in capturing the internal EPR by accounting for irreversibility associated with the directionality of $P( \ell_\beta|\ell_\alpha)$. 
Indeed, by applying the same coarse-graining procedure as in the square lattice (\fref{fig:2}B), we observe a power law decay of $\sigma_2^\ell$ with resolution (quantified by $n_{\rm{E}}$).
The exponent is $\sim 0.3$, much smaller than the exponent for $\sigma_1$ ($\sim 0.6$). 
Again, the combined estimator $\sigma_2=\sigma_2^\ell+\sigma_2^t$ shows similar scaling (Appendix B) since it is dominated by $\sigma_2^\ell$.
The power law persists until $n_{\rm{B}}$ reaches the size of the limit cycle, which is $n_B \sim 100$ for the parameters used in \fref{fig:2}B. 
For smaller resolutions, the cycle is not visible and the captured EPR will enter a different scaling regime.

In this model, we also compute $\sigma_2^t$ for different volumes $V$, which tunes both the size and the period of the limit cycle. As shown in \fref{fig:2}C, $\sigma_2^t$ is non-monotonic in $n_{\rm{B}}$, with the peak position $n_{\rm{B,m}}$ (dashed lines) increasing with $V$. 
In order to compare the peak position with the period of the limit cycle, we convert $n_{\rm{B}}$ to the characteristic timescale defined as the inverse dynamic activity, $\tau\equiv \mathcal{K}_{\rm{obs}}^{-1}$.
$\tau$ represents the average waiting time between transitions at each coarse-grained level and $\mathcal{K}_{\rm{obs}}$ is the average measurement frequency.
This converts $\sigma_2^t$ to a function of $\tau$ (\fref{fig:2}D), whose peak position $\tau^\star$ is approximately linear with the oscillation period (inset). In other words, when the resolution is tuned to get maximally asymmetric waiting time distributions, the resulting coarse-grained timescale is related to the oscillation period, which is an internal dissipative timescale.

Notably, a similar non-monotonic relation between the EPR and the coarse-grained timescale has been recently reported in the actomyosin cortex of starfish oocytes~\cite{tan_scale-dependent_2021}. In these experiments, irreversibility was measured by the KL divergence of the displacement of cortical granules after different lag times: $r^{\vartheta}=r(t+{\vartheta})-r(t)$ versus the displacement in the time reverse process. 
The lag time $\vartheta$ is a timescale for coarse-graining because it quantifies the frequency of measurements. It is analogous to the coarse-grained timescale $\tau$ defined here. 
It was found that the irreversibility $\sigma$ is non-monotonic in $\vartheta$, with its peak position $\vartheta^\star$ being a characteristic timescale.
Strikingly, when the oscillation frequency of Rho-GTP patterns was tuned by modulating the intracellular ATP concentration, the characteristic timescale  $\vartheta^\star$ was approximately linear with the period of Rho-GTP concentration oscillations (figure 3D in \cite{tan_scale-dependent_2021}), consistent with the linear relation in the Brusselator (\fref{fig:2}D, inset). 
Therefore, our results provide a potential theoretical explanation for the observed scale-dependent irreversibility and may afford broader applicability to a variety of nonequilibrium living systems.
\\

\subsection{Two-component Schl\"{o}gl model}\label{subsec:schlogl}

The crossover of distinct scaling regimes, indicating dissipative scales, is absent in the square lattice with random rates and not particularly evident for the Brusselator. For this reason, we consider the two-component Schl\"{o}gl model~\cite{schlogl_chemical_1972, vellela_stochastic_2009}. The system has two compartments with identical chemical reactions. The particles can either undergo reactions within compartments or hop between compartments. Thus, the model is analogous to the two-site active Ising model~\cite{yu_flocking_2022}. Let $Z=X,Y$ be the particles in the two compartments. The reactions within each compartment are:
\begin{equation}
    B \underset{k_{0}}{\stackrel{k_{2}}{\rightleftharpoons}} Z,
    \qquad 2Z+A \underset{k_{1}}{\stackrel{k_{3}}{\rightleftharpoons}} 3Z,
\end{equation}
with identical rates in the two compartments. The concentrations $[A]$ and $[B]$ are fixed by chemostats. The compartments exchange particles with rate $\gamma$: $X \underset{\gamma}{\stackrel{\gamma}{\rightleftharpoons}} Y$. 

\begin{figure}[t]
    \centering
    \includegraphics[width=0.9\textwidth]{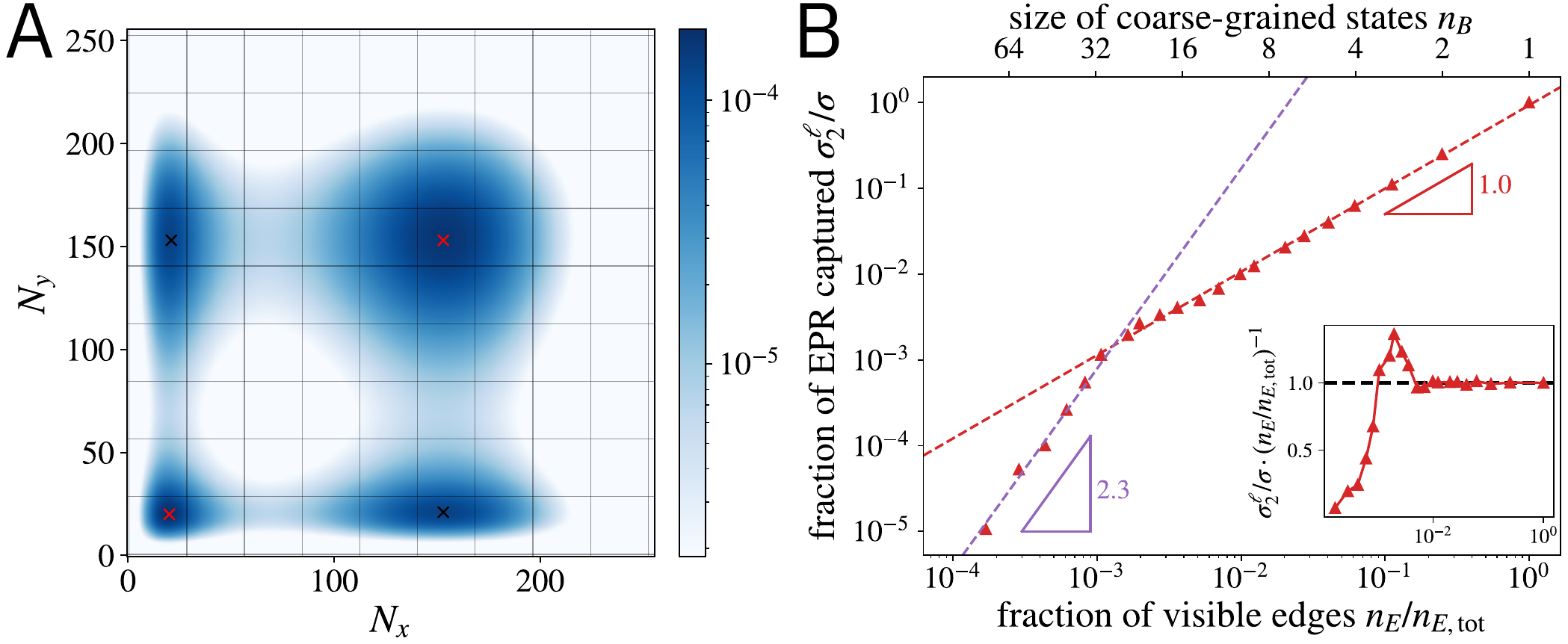}
    \caption{Two-compartment Schl\"ogl model.
        (A) Steady-state probability distribution, with grids indicating the transition scale $n_{\rm{B}}^\star$. Red and black crosses indicate local maxima.
        (B) Fraction of EPR captured $\sigma_2^\ell/\sigma$  shows two scaling regimes (red and purple dashed lines). They cross at the transition scale $n_{\rm{B}}^\star$. Inset: $\sigma_2^\ell$ normalized by the number of visible transitions.
        Parameters: $k_0=k_3=1$, $k_1=k_2=4$, $\gamma=0.01$, $V=60$.}
    \label{fig:3}
\end{figure}

At intermediate $\gamma$, the system has four locally stable (macroscopic) states [see \fref{fig:3}A for the probability distribution in the $(N_x, N_y)$ plane] representing homogeneous/inhomogeneous high/low-density states. 
Each of the four states is dissipative since the inter-compartment exchange does not commute with reactions within the compartments, resulting in local vortices. The four states are also connected by large-scale dissipative flows that form global vortices. 
Thus, we expect $\sigma_2^\ell$ to exhibit distinct scaling exponents for local and global flows. At high resolution, coarse-graining reduces dissipation predominantly by operating on local flows, while at low resolution it does so by lumping global (large-scale) flows between the four states. 
Indeed, \fref{fig:3}B shows two scaling regimes: at large (small) block size ($n_{\rm{B}}$), $\sigma_2^\ell$ decreases much faster (slower). 
The two regimes cross at $n_{\rm{B}}^\star\approx 28$, which is approximately the spread of each macroscopic state (grids in \fref{fig:3}A). 
The scaling exponent is larger at low resolutions (larger $n_{\rm{B}}$) because coarse-graining only operates on the transitions between stable states while ignoring their internal structures. 
Thus, the crossover of scaling regimes reveals the characteristic scales of the dissipative dynamics.
A similar crossover can be identified in experimental data as long as the measurement is at a resolution higher than $n_{\rm{B}}^\star$.
\\

%
%

\subsection{Networks with non-regular topologies}\label{subsec:networks}

\begin{figure*}[t]
    \centering
    \includegraphics[width=.8\textwidth]{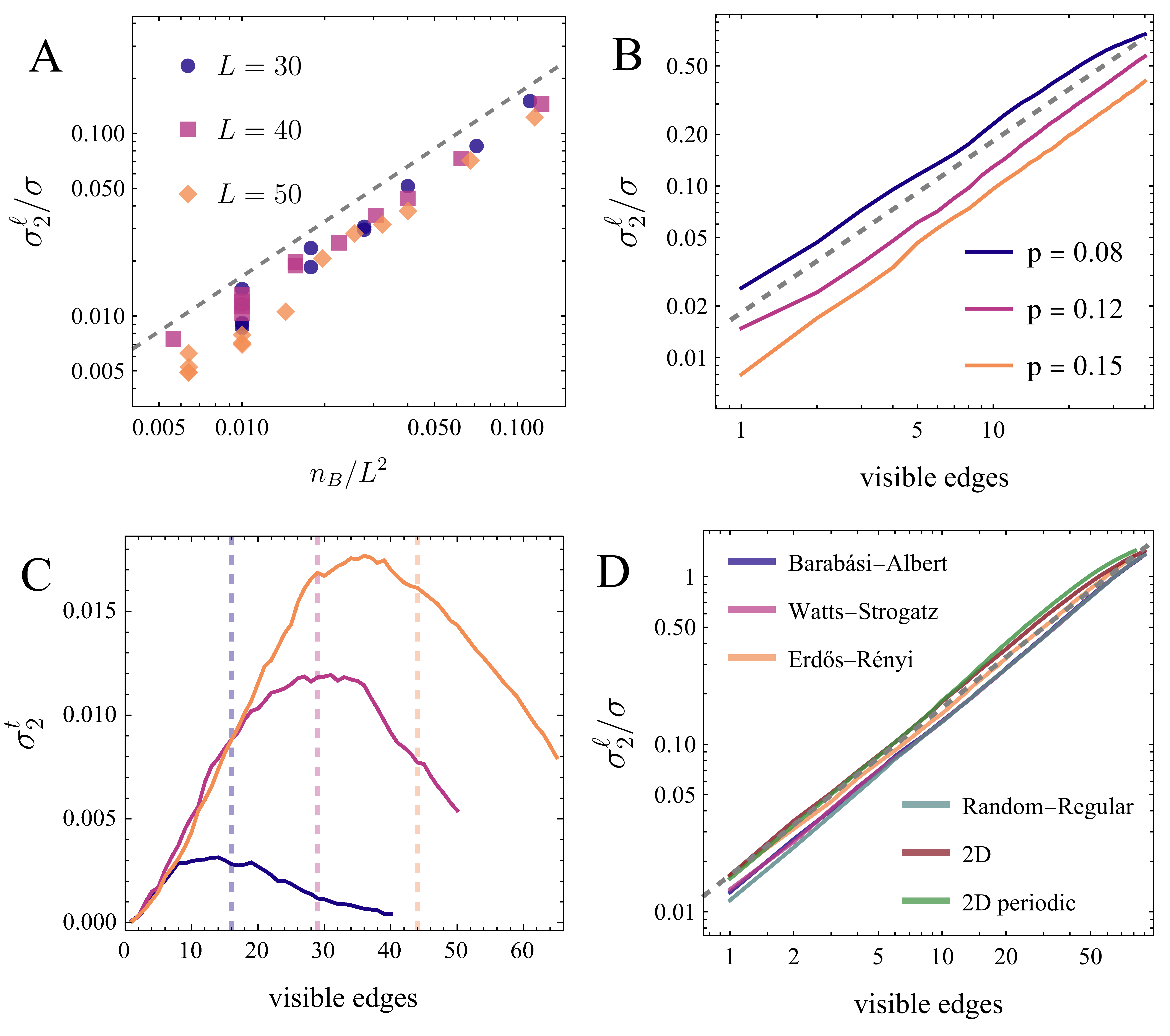}
    \caption{(A) Fraction of EPR captured $\sigma_2^\ell/\sigma$ for scale-free networks embedded in a $L\times L$ lattice.
    The dashed line has slope $1$.
    Embedded networks have degree distribution $P(k) \propto k^{-3}$, minimum degree $k_{\rm{min}} = 3$, and maximum connection distance $10 \sqrt{k}$. 
    (B) Fraction of EPR captured $\sigma_2^\ell/\sigma$ for Erd\H{o}s-R\'enyi networks, averaged over random selections of visible edges. The dashed line has slope $1$. Networks have $N=40$ vertices and varying parameter $p$. 
    (C) $\sigma_2^t$ for networks in (B); vertical dashed lines mark the respective number of cycles divided by 2. 
    (D) Log-log plot of $\sigma_2^\ell/\sigma$ with a dashed line of slope 1; curves represent the average over 400 realizations of randomly selecting edges in networks of distinct types, all with $N=60$ vertices; further parameters: $m=3$ (Barabási-Albert), $m=6$ and $p=0.7$ (Watts-Strogatz), $p=0.1$ (Erd\H{o}s-R\'enyi), $d=6$ (2D Random-Regular).
    }
    \label{fig:4}
\end{figure*}

The previous arguments and illustrations cover networks with regular structures, but the main findings are also witnessed in network of complext topologies, suggesting an even broader applicability.
For direct lumping, previous work has shown that the EPR scaling only emerges in networks with a self-similar structure, such as a scale-free network~\cite{yu_inverse_2021}. Here we find the scaling of $\sigma_2^\ell$ in lattice-embedded scale-free networks~\cite{rozenfeld_scale-free_2002, kim_geographical_2004} with an exponent of approximately $1$ (\fref{fig:4}A), an improvement from the direct coarse-graining approach~\cite{yu_inverse_2021}.

In contrast to lattice-embedded networks, where the measurement resolution can be captured by the size of blocks used in coarse-graining, many real networks have no apparent embedding, which makes it difficult to define the coarse-graining procedure. 
For these systems, we describe resolution by assuming that only a random subset of transitions are visible. 
Although these edges do not necessarily divide the network into equal-sized coarse-grained states, they provide an apparent measure of the irreversibility of the coarse-grained dynamics. We study how $\sigma_2^\ell$ and $\sigma_2^t$ scale with the number of visible transitions. 

We first consider the Erd\H{o}s-Rényi model, a canonical model for real-world networks with random topology. It is characterized by two values, the number of vertices and the probability $p$. The network is constructed by starting with the vertices only, and to each pair of vertices an edge is created with probability $p$. Despite the topology being random, the interplay between the two parameters leads to rich phenomena, such as the emergence of a giant connected component through a phase transition \cite{erdosEVOLUTIONRANDOMGRAPHS}. Here, we consider Erd\H{o}s-Rényi networks that after the generation turn out to be irreducible, so the system reaches a stationary state, and randomized transition rates are assigned to both directions of each edge.
By growing the subset of visible edges through random selection, we find an increasing relation between $\sigma_2^\ell$ and the number of visible edges. 
When averaged over the order of edge selection, $\sigma_2^\ell$ exhibits robust scaling with the number of visible edges with an exponent of approximately $1$ (\fref{fig:4}B).
The same procedure reveals that $\sigma_2^t$ varies non-monotonically (\fref{fig:4}C), consistent with our intuition from the Brusselator model (Section~\ref{subsec:brusselator}). The peaks of $\sigma_2^t$ are associated with an internal notion of dissipative scale, and it remains unclear how to determine it through the topological or dynamical properties of the system since there is no unique way of defining the complexity of a network. As a candidate, we compare the peak of $\sigma_2^t$ with half the number of cycles in the network (dashed lines of \fref{fig:4}C), which are the elementary units of dissipative fluxes in nonequilibrium reaction networks~\cite{schnakenberg_network_1976,zia_probability_2007}. This comparison is also motivated by the fact that a visible edge per cycle causes $\sigma_2^t$ to vanish and $\sigma_2^\ell$ to capture the full EPR \cite{van_der_meer_thermodynamic_2022}. We observe that the peaks are roughly related to the number of cycles, which is a factor in defining the internal dissipative scale but not the sole ingredient.

A plethora of models that generate random topologies are relevant in the study of real-world networks \cite{dorogovtsev2003evolution}, and properties such as degree distribution and number of cycles can be substantially different among them. To study the effect of network topology, we perform the same analysis in other remarkable networks: Barabási-Albert, Watts-Strogatz, random-regular, and a two-dimensional grid graph with and without boundary conditions. We observe that the scaling behavior of the apparent EPR measured by $\sigma_2^\ell$ is robust with respect to topology (\fref{fig:4}D), presenting a very similar exponent of $\sim 1$. It is striking that these results are robust to the state-space structure as well as to the specific coarse-graining procedure, suggesting the general applicability of this approach to a broad class of systems.

%
%

\section{Discussion}\label{sec:discussion}

Our results demonstrate that both the sequence of transitions and the distribution of intertransition times can help extract the entropy production rate from measurements with limited resolutions. In all cases studied here, ranging from chemical reaction systems to networks with complex topologies, $\sigma_2^\ell$ scales with resolution following power laws.
The scaling exponent is smaller than that of the less-informed estimator $\sigma_1$, often close to unity, representing a linear scaling with the number of visible transitions. It would be revealing to explore whether higher-order estimators have even smaller exponents. 
Although the scaling of $\sigma_2^\ell$ can be conceptually understood by generalizing the scaling argument~\cite{yu_inverse_2021} for $\sigma_1$ from state space to transition space, it remains unclear how the scaling exponent for $\sigma_2^\ell$ can be determined quantitatively. 
Since the exponent for $\sigma_1$ depends on the network structure and flux correlations, we hypothesize that the exponent for $\sigma_2^\ell$ is related to correlations in transition space, which may be determined through a renormalization group analysis~\cite{yu_state-space_2022}. 
It will be interesting to investigate whether the exponent uncovers more properties of the underlying dissipative dynamics and how it depends on the physical properties of the system. 

Regarding the detection of internal scales, the crossover between multiple scaling regimes of $\sigma_2^\ell$ provides a way to detect characteristic length scales in a dissipative system. On the other hand, the intertransition-time-based estimator $\sigma_2^t$ varies non-monotonically with resolution, with its peak reflecting an internal dissipative scale at which the waiting time distributions become maximally asymmetric. 
Our results show that the $\sigma_2^t$ peak is related to the period of oscillation in the Brusselator and to the number of cycles in a complex network. However, further studies are needed to elucidate how the network topology and transition dynamics affect the crossover in $\sigma_2^\ell$ and the non-monotonicity of $\sigma_2^t$. In addition, the power laws with distinct exponents separated by a kink in $\sigma_2^\ell$ are reminiscent of the (inverse) energy cascades in turbulent flows~\cite{alexakisCascadesTransitionsTurbulent2018, dewitPatternFormationTurbulent2024,yu_inverse_2021}, where energy is injected at a given scale and separates the cascade into distinct regimes; this similarity could lead to additional insights into dissipative scales.

Both $\sigma_2^\ell$ and $\sigma_2^t$ can be readily computed for experimental data: while $\sigma_2^\ell$ can be estimated directly from a histogram of visible transitions, $\sigma_2^t$ requires evaluating the Kullback-Leibler divergence from a finite data set of continuous random variables, for example, with the algorithm in Refs.~\cite{4595271, Harunari_KLD_estimation}.
In addition to estimating the irreversibility $\sigma_2^\ell+\sigma_2^t$, one can also introduce more coarse-graining levels at post-processing to investigate the dependence of $\sigma_2^\ell$ and $\sigma_2^t$ on resolution. Therefore, these behaviors can be used as a tool to uncover hidden properties. It may be fruitful to combine the approach with experiments in spatially extended dissipative systems such as the actomyosin cortex~\cite{tan_scale-dependent_2021} and microtubule active gels~\cite{foster_dissipation_2023} to reveal more information on the irreversibility across length scales. 

Remarkably, the predicted non-monotonic relation between irreversibility and resolution has been observed experimentally in the actomyosin cortex~\cite{tan_scale-dependent_2021}. There, the dissipative timescale was identified by the lag time corresponding to maximum irreversibility, which is analogous to the coarse-graining scale corresponding to maximum $\sigma_2^t$ proposed in this work (\fref{fig:2}D). This speaks to the applicability of our formalism to experimental measurements, especially in active matter systems. 
From a theoretical perspective, however, there is a slight nuance between transition-based measures and stroboscopic observations, namely snapshots separated by different lag times as were used in ref.~\cite{tan_scale-dependent_2021}. A fruitful future direction is to build a rigorous connection between the two approaches.

In a given model, the dissipative timescale and length scale might be connected by simple relations, but universal considerations cannot be drawn at this point and deserve further investigation. 
We also highlight that, depending on the system, the resolution in time and space are of different relevance. For instance, in the two-compartment Schl\"ogl model, a lower spatial resolution might completely miss the dissipative dynamics inside each metastable state, whereas a lower time resolution might still be able to capture it provided the typical escape time from metastable states is sufficiently long. Understanding the interplay between both notions of resolution can guide optimal strategies for experimental applications.

Our empirical results show that at very low resolutions, $\sigma_2^\ell$ tends to be much smaller than the true EPR due to the power-law scaling, highlighting that a route to estimate the dissipated energy should include either resolution enhancement or methods that rely on additional information beyond the pairwise statistics of transitions. It will also be interesting to investigate alternative estimators that also account for a priori knowledge on the nature of the hidden structures, such as the possible chemical reactions or the topological state-space structure, these may provide more information on the underlying dissipative dynamics. 

%
%

\ack

We thank Dr.~Junang Li for useful discussion and comments on an early version of the manuscript. 
Q.~Y. thanks Prof.~Tzer Han Tan for stimulating discussions on the irreversibility measurements in the actomyosin cortex. The work by Q.~Y. was supported in part by a Harold W. Dodds Fellowship from Princeton University.
The work by P.~H. was supported by the project INTER/FNRS/20/15074473 funded by F.R.S.-FNRS (Belgium) and FNR (Luxembourg). 
This work was initiated at a Physics of Life symposium at the Center for the Physics of Biological Function (NSF PHY-1734030).

%
%

\appendix
\section*{Appendix A: Survival matrix technique}\label{app:survival}
\setcounter{section}{1}

To compute $\sigma_2^\ell$ and $\sigma_2^t$ for the coarse-grained system, we use the survival matrix method \cite{harunari_what_2022} to analytically derive the joint probabilities $P(\ell_\alpha, \ell_\beta) =  P(\ell_\beta \vert \ell_\alpha) P(\ell_\alpha)$ and the waiting time distributions $P(t|\ell_\alpha, \ell_\beta)$. This amounts to solving a first-passage problem between given transitions. 

We consider a continuous-time Markov chain whose dynamics is described the rate matrix $\mathbf{R}$ through the master equation $d_t \boldsymbol{p}_t = \mathbf{R} \boldsymbol{p}_t$.
The off-diagonal element $[\mathbf{R}]_{ij}$ is the transition rate from state $j$ to $i$ ($\neq j$).
The diagonal element $[\mathbf{R}]_{ii} = -\sum_{j\neq i} [\mathbf{R}]_{ji}$ is the escape rate of leaving state $i$. 
The goal is to compute the first-passage time distribution between a subset of transitions that are visible.
To this end, we introduce the survival matrix $\mathbf{S}$, which is defined by removing all visible transitions in the off-diagonal elements from the rate matrix $\mathbf{R}$ while preserving diagonal elements. The survival matrix captures the internal evolution of the system between observable transitions. To be more precise, $[\exp (\mathbf{S}t)]_{j,i}$ is the probability of being in the microscopic state $j$ at time $t$ after starting at microscopic state $i$ at time $0$, without taking any visible transitions.

In this work, only the transitions between coarse-grained states are visible. Here, we compute the first-passage probabilities between the microscopic transitions $\ell_{\alpha, i}$ and $\ell_{\beta, j}$ that connect distinct coarse-grained states. 
They can be used to derive the probabilities of coarse-grained transitions through Eqs.~\eref{eq:cg_joint} and \eref{eq:cg_marginal}. 
The visible transitions divide the system into distinct coarse-grained states, which allows applying the survival matrix approach to each coarse-grained state individually. 
For each coarse-grained state, the survival matrix reads
\begin{eqnarray}
    [\mathbf{S}]_{ij} = [\mathbf{R}]_{ij} - \delta_{ij} \sum_{k} [\mathbf{R}]_{kj},
\end{eqnarray}
where $i,j$ enumerate all microscopic states inside the coarse-grained state, while $k$ runs over states both inside and outside the coarse-grained state. While the off-diagonal terms capture all the internal transitions, the diagonal ones include both internal and external (i.e. escaping) transitions. 

Let $\mathsf{s}(\ell)$ and $\mathsf{t}(\ell)$ be the source and target states of transition $\ell$, respectively. The joint probability of a transition $\ell_{\beta, j}$ and intertransition time $t$ conditioned on the previous transition $\ell_{\alpha, i}$ is given by
\begin{equation}\label{eq:firsttransitionproblem}
    P( \ell_{\beta,j}, t \mid \ell_{\alpha,i}) = [\mathbf{R}]_{ \mathsf{t} ( \ell_{\beta,j} ) , \mathsf{s} ( \ell_{\beta,j} ) } [ e^{\mathbf{S} t }]_{ \mathsf{s} ( \ell_{\beta,j} ) , \mathsf{t} ( \ell_{\alpha,i} ) },
\end{equation}
which can be marginalized to obtain the conditional transition probability
\begin{equation}
    P( \ell_{\beta,j} \mid \ell_{\alpha,i}) = - [\mathbf{R}]_{ \mathsf{t} ( \ell_{\beta,j} ) , \mathsf{s} ( \ell_{\beta,j} ) } [ \mathbf{S}^{-1}]_{ \mathsf{s} ( \ell_{\beta,j} ) , \mathsf{t} ( \ell_{\alpha,i} ) }.
\end{equation}
The ratio gives the probability of the intertransition time between two transitions
\begin{equation}
    P( t \mid \ell_{\alpha,i}, \ell_{\beta,j}) = \frac{P( \ell_{\beta,j}, t \mid \ell_{\alpha,i})}{P( \ell_{\beta,j} \mid \ell_{\alpha,i})}.
\end{equation}
Lastly, it is also necessary to compute the absolute probability of a single transition
\begin{equation}
    P(\ell) = \frac{ [\mathbf{R}]_{ \mathsf{t} ( \ell) , \mathsf{s} ( \ell ) } p_{\mathsf{s} ( \ell )}  }{ \sum_{\ell'} [\mathbf{R}]_{ \mathsf{t} ( \ell') , \mathsf{s} ( \ell' ) } p_{\mathsf{s} ( \ell' )} }.
\end{equation}

In the present work, we consider that transitions between coarse-grained states $\ell_\alpha$ cannot be resolved as $\ell_{\alpha,i}$, hence we combine the probabilities above to obtain the quantities involved in $\sigma_2^\ell$ and $\sigma_2^t$, as described by Eqs.~\eref{eq:cg_joint} and \eref{eq:cg_marginal}.

\begin{figure*}[tb]
    \centering
    \includegraphics[width=0.8\textwidth]{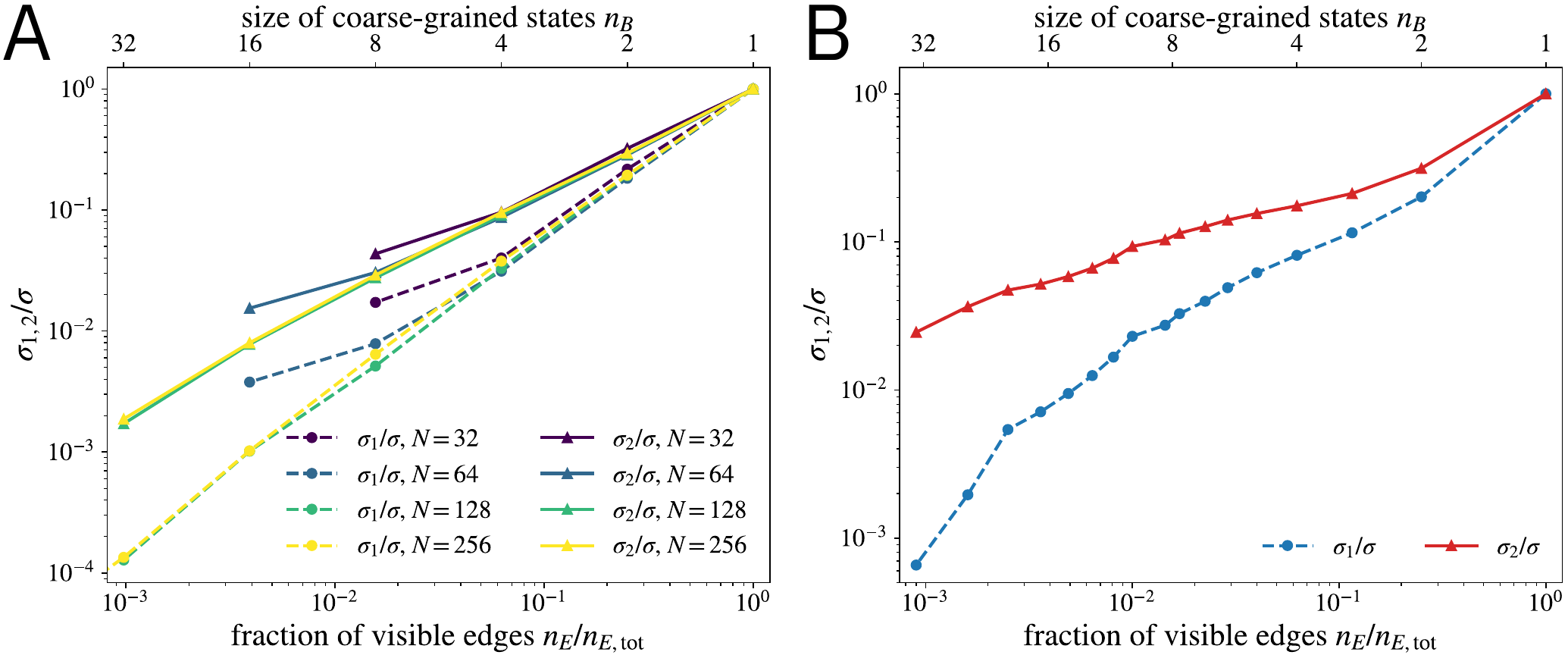}
    \caption{
        Comparing the one-step estimator $\sigma_1$ and two-step estimator $\sigma_2=\sigma_2^\ell+\sigma_2^t$ in square lattice (A) and Brusselator (B). 
        The square lattice uses i.i.d. transition rates as was used in \fref{fig:1}.
        The Brusselator model uses the same parameters as in \fref{fig:2} with volume $V=6$.
    }
    \label{fig:s1}
\end{figure*}

\section*{Appendix B: Comparing $\sigma_1$ and $\sigma_2$}\label{app:compare}
\setcounter{section}{2}

In the main text, we have analyzed $\sigma_2^\ell$ and  $\sigma_2^t$ separately to unravel their own properties. Here we compare the combined two-step estimator $\sigma_2=\sigma_2^\ell+\sigma_2^t$ with the one-step estimator $\sigma_1$. As shown in \fref{fig:s1}, compared to $\sigma_1$, $\sigma_2$ preserves a considerably larger fraction of the entropy production rate as the system is coarse-grained. 
Moreover, panel A shows that increasing the size of the square lattice expands the range for power-law scaling and that deviation from power law occurs only when $n_B$ becomes comparable to the system size.

\clearpage
\section*{References}

\bibliographystyle{unsrt}

\bibliography{epr_scaling}

\end{document}